\documentclass[aps,twocolumn,10pt,longbibliography,superscriptaddress]{revtex4-1}

\usepackage{amsmath}
\usepackage{amsfonts}
\usepackage{latexsym}
\usepackage{graphicx}
\usepackage{color}
\usepackage{bm}
\usepackage{float}

\definecolor{Blue}{rgb}{0,0,1}

\newcommand{\bra}[1]{{\left\langle{#1}\right\vert}}
\newcommand{\ket}[1]{{\left\vert{#1}\right\rangle}}
\newcommand{\tr}{\operatorname{tr}}

\usepackage{hyperref} %Hyrperref package: only for revtex version
\usepackage[all]{hypcap}
\hypersetup{pdfauthor = {First Author},pdftitle = {Experimental implementation of superposition of causal order}}

\begin{document}
\title{Experimental Superposition of Orders of Quantum Gates}
\author{Lorenzo M. Procopio}
\thanks{Corresponding author: lorenzo.procopio@univie.ac.at}
\affiliation{Faculty of Physics, University of Vienna, Boltzmanngasse 5, A-1090 Vienna, Austria}

\author{Amir Moqanaki}
\affiliation{Faculty of Physics, University of Vienna, Boltzmanngasse 5, A-1090 Vienna, Austria}

\author{Mateus Ara{\'u}jo}
\affiliation{Faculty of Physics, University of Vienna, Boltzmanngasse 5, A-1090 Vienna, Austria}
\affiliation{Institute for Quantum Optics and Quantum Information, Austrian Academy of Sciences, Boltzmanngasse 3, A-1090 Vienna, Austria}
\author{Fabio Costa}
\thanks{Present address: Centre for Engineered Quantum Systems, School of Mathematics and Physics, The University of Queensland, St Lucia, QLD 4072, Australia}
\affiliation{Faculty of Physics, University of Vienna, Boltzmanngasse 5, A-1090 Vienna, Austria}
\affiliation{Institute for Quantum Optics and Quantum Information, Austrian Academy of Sciences, Boltzmanngasse 3, A-1090 Vienna, Austria}

\author{Irati A. Calafell}
\affiliation{Faculty of Physics, University of Vienna, Boltzmanngasse 5, A-1090 Vienna, Austria}

\author{Emma G. Dowd}
\thanks{Present address: Department of Physics, Harvard University, Cambridge, MA}
\affiliation{Faculty of Physics, University of Vienna, Boltzmanngasse 5, A-1090 Vienna, Austria}

\author{Deny R. Hamel}
\thanks{Present address: D\'{e}partement de physique et d'astronomie, Universit\'{e} de Moncton, Moncton, New Brunswick E1A 3E9, Canada}
\affiliation{Faculty of Physics, University of Vienna, Boltzmanngasse 5, A-1090 Vienna, Austria}

\author{Lee A. Rozema}
\affiliation{Faculty of Physics, University of Vienna, Boltzmanngasse 5, A-1090 Vienna, Austria}
% new line for the theorists?
\author{\v{C}aslav Brukner}
\affiliation{Faculty of Physics, University of Vienna, Boltzmanngasse 5, A-1090 Vienna, Austria}
\affiliation{Institute for Quantum Optics and Quantum Information, Austrian Academy of Sciences, Boltzmanngasse 3, A-1090 Vienna, Austria}

\author{Philip Walther}
\thanks{Corresponding author: philip.walther@univie.ac.at}
\affiliation{Faculty of Physics, University of Vienna, Boltzmanngasse 5, A-1090 Vienna, Austria}

%%%%%%%%%%%%%%%%%%%%%%%%%%%%%%%%%%%%%%%%%%%%%%%%%%%%%%%%%%%%%%%%%%%%%%%%%%%%%%%%%%%%%%%%%%%%%%%%%%%%%%%%%%%%%

\begin{abstract}
In a quantum computer, creating superpositions of quantum bits (qubits) in different {states} can lead to a speed-up over classical computers \cite{Deutsch1989}, but quantum mechanics also allows for the superposition of 
quantum circuits \cite{Chiribella2013}.
In fact, it has recently been theoretically predicted that superimposing quantum circuits, each  { with a}  different gate order, could provide quantum computers {with} an even further computational advantage \cite{Chiribella2012,Araujo2014,brukner2014}. %Colnaghi2012
Here, { we experimentally demonstrate this enhancement by applying two quantum gates in a superposition of both possible orders  to determine whether the two gates commute or anti-commute.} We are able to make this determination with only {\textit{a single}}  use (or query) of each gate, while all quantum circuits with a fixed order of gates would require at least two uses of one of the gates \cite{Chiribella2012}.
Remarkably, when the problem is scaled to $N$ gates, creating a superposition of quantum circuits is likely to provide an exponential advantage over classical algorithms, and a { linear } advantage over  quantum algorithms with fixed gate order \cite{Araujo2014}.
The new resource that we exploit in our experiment can be interpreted as a  ``superposition of causal orders''.
We demonstrate such a superposition could allow some quantum algorithms to be implemented with an efficiency that is unlikely to be achieved on a quantum computer with a fixed gate order.
\end{abstract}

\maketitle %This is for the revtex format, which requires \maketitle to come after the abstract. Comment it out for nature format (it is already included in the Nature preamble).

One of the most useful methods for quantifying the performance of a quantum algorithm is its query complexity.
Loosely speaking, this is the number of times that a quantum gate is used (or queried).
The use of the query complexity is motivated by the assumption that applying a gate is a cost that we wish to minimize.
In an optical quantum computer this cost would be either another physical copy of the gate (say a different set of waveplates, or interferometer), or a repeated usage of the same gate at a later time.
On the other hand, in an ion-trap \cite{cirac1995} or super-conducting \cite{dicarlo2009} quantum computer the cost would be the application of another pulse sequence to the qubits.
Given that one of the main difficulties in creating a scalable quantum computer is the implementation of multiple gates, techniques to reduce the query complexity are essential for practical quantum computing.

\begin{figure}
\begin{center} 
\scalebox{.63}{\includegraphics{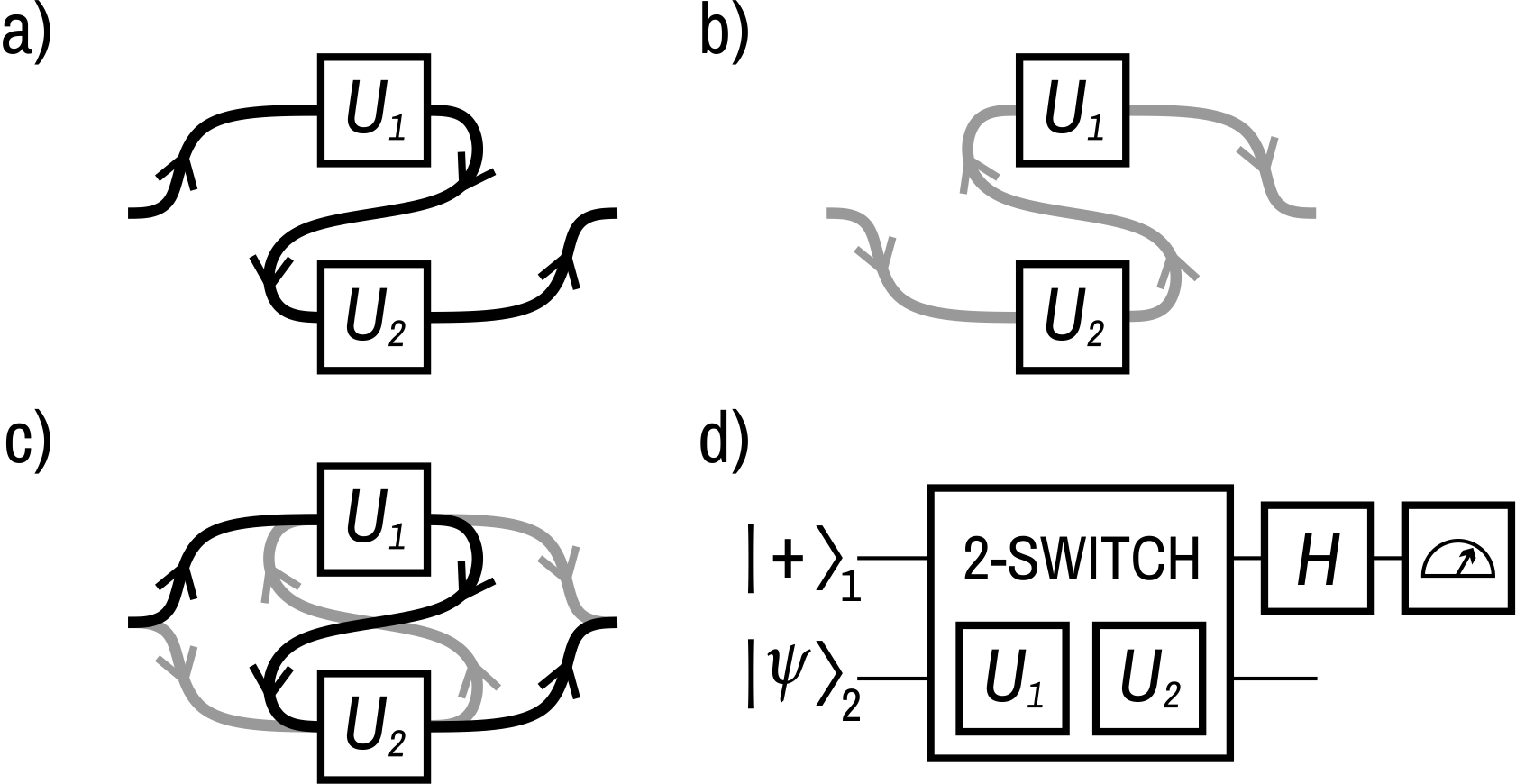}}
\caption{\footnotesize a) Given two unitary gates, $U_1$ and $U_2$, the circuit model allows us to wire them { in}  one of two {possible } ways: either $U_1$ before $U_2$, or b) $U_2$ before $U_1$. c) Quantum mechanics allows us { to} coherently control both options, such that the qubit sees both $U_1$ before $U_2$, \textit{and} $U_2$ before $U_1$. 
d) The 2-SWITCH operation applies $U_1$ and $U_2$ to qubit 2 in \textit{both orders}, as shown in panel c, dependent on the state of qubit 1. 
Unless {at least one of } $U_1$ and $U_2$ { is} used more than once, the 2-SWITCH operation \textit{cannot} be implemented with standard circuit-model elements.
To be explicit, the 2-SWITCH applies $U_1 U_2$ to $\ket{\psi}_2$ (the lower qubit) if the upper qubit is in $\ket{0}_1$, and $U_2 U_1$ to $\ket{\psi}_2$ if the upper qubit is in $\ket{1}_1$. 
Measuring the state of qubit 1 in the $\ket{\pm}$ basis allows one to unambiguously decide if $U_1$ and $U_2$ commute or anti-commute with only a single use of each gate.
In this circuit, $H$ is the Hadamard gate, and $\ket{\pm} = (\ket{0}\pm\ket{1})/\sqrt{2}$. } \label{cartoon}
\end{center}
\end{figure}

\begin{figure*}[ht]
\begin{center} 
\includegraphics[width = .95\textwidth]{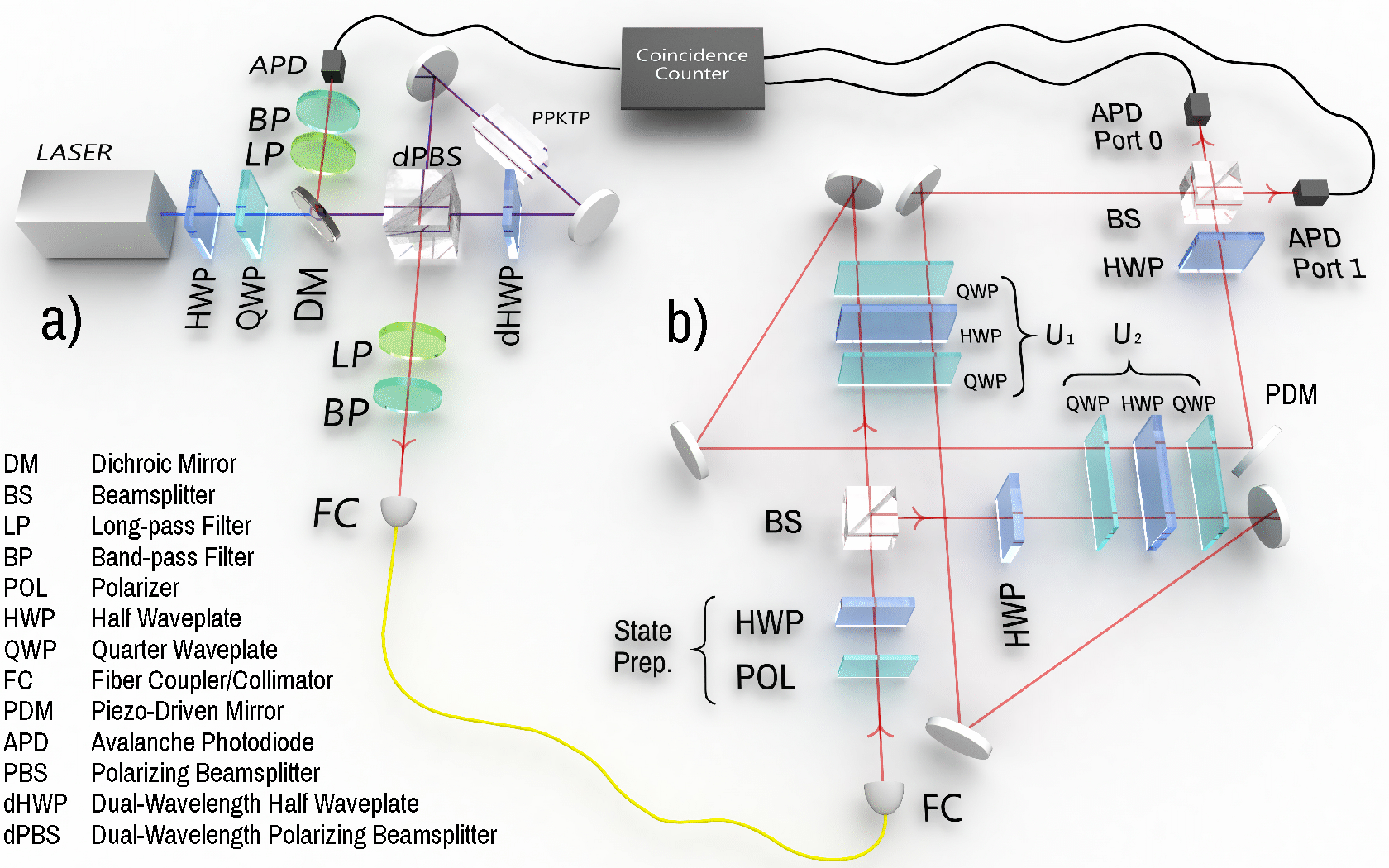}
\caption{\footnotesize Our optical implementation to distinguish whether a pair of unitary gates commute or anti-commute with only a single copy of each gate.
The photons for our experiment are generated in a separable polarization state using a Sagnac source (a).
One photon is used as a herald, and the second is fed into the interferometer (b).
The unitary gates in question are each implemented with three waveplates, and act on the polarization of single photons.
} \label{expt}
\end{center}
\end{figure*}

Just as in a classical electronic circuit, in a fixed-order quantum circuit one connects a series of logic gates by wires (see Fig. 1a or Fig. 1b).
This is an intuitive and extremely powerful method for designing quantum algorithms, but there are advantages to other models.
One example is measurement-based quantum computing \cite{raussendorf2001}, a different paradigm than the circuit model which paved the way for many experimental implementations of quantum algorithms \cite{walther2005,prevedel2007}.
Unlike other models of computation, what is considered here is a strict extension of the quantum circuit model, which therefore allows for additional computational power.
The particular extension we study in our experiment is to allow for superpositions of different quantum circuits; i.e. to coherently control which quantum circuit is applied on an input state (see Fig. 1c).
In this case, the order of quantum gates acting on a set of qubits could be controlled by the state of another set of qubits -- this is not allowed in the standard quantum circuit model, wherein the gate order is independent of the state of the qubits \cite{Chiribella2013}.

Coherently controlling the order of quantum gates conditioned on the state of a set { of} qubits is a new type of operation.
A proposal for one such operation is the ``N-SWITCH'', which takes $N$ different gates and applies them in a given superposition { of }different permutations \cite{Colnaghi2012}.
Using this operation, { a quantum algorithm has recently been proposed to solve a specific problem with a query complexity of $O(N)$, while a fixed-order circuit is likely to require  $O(N^2)$ queries to solve the same problem \cite{Araujo2014,brukner2014}.}

The problem we study in this experiment is a specific case of a computational problem proposed in Ref. 4, based on Ref. 3.
One is presented with two unitary gates, $U_1$ and $U_2$, and the guarantee that $U_1$ and $U_2$ either commute or anti-commute (but $U_1$ and $U_2$ are otherwise unknown and arbitrary).
The goal is to determine which statement is true.
In the standard circuit model this cannot be done with a single use of each gate \cite{Chiribella2012}; this limitation is evident in all previous experimental investigations of commutation relations, which all used one gate at least twice \cite{yao2010,kim2010}.
However, using the 2-SWITCH operation to apply $U_1$ and $U_2$ in a superposition of both orders allows us to distinguish whether the gates commute or anti-commute with only one use of each gate.
To see this, consider using the 2-SWITCH operation in the circuit shown in Fig. 1d.
Given that the 2-SWITCH applies $U_1U_2$ if the upper qubit is in $\ket{0}_1$ and $U_2U_1$ if upper qubit is in $\ket{1}_1$, it is straightforward to show that, when the input to the circuit is initially in the state $\frac{1}{\sqrt{2}}(\ket{0}_1+\ket{1}_1)\ket{\psi}_2$ (where $\ket{\psi}_2$ is an arbitrary state of qubit 2), the result is the state
\begin{equation}
\frac{1}{2}\ket{0}_1\{U_1,U_2\}\ket{\psi}_2 + \frac{1}{2}\ket{1}_1[U_1,U_2]\ket{\psi}_2,
\end{equation}
where $[U_1,U_2]$ and $\{U_1,U_2\}$ are the commutator and anti-commutator of $U_1$ and $U_2$, respectively.
{Given the guarantee of either commutation or anti-commutation, 
if the upper qubit is measured and found in $\ket{0}_1$ we know for certain that the gates commute;
on the other hand, if it is found in $\ket{1}_1$ we know for certain that the gates anti-commute.
Thus one can unambiguously distinguish between the two cases.}
Note that, although Fig. 1d shows the 2-SWITCH operation { as a gate} in a quantum circuit, { it cannot be implemented by querying $U_1$ and $U_2$ only once in a fixed order.}

Although creating superpositions of circuits is a conceptually simple idea, it is not immediately clear how it could be done { in} the laboratory.
The most obvious solution is to place the physical circuit elements -- such as wires or optical fibres connecting the gates -- in a quantum superposition.
However, this would require quantum control over macroscopic systems, and is likely to remain unattainable in the foreseeable future.
{Instead, we use internal degrees of freedom of our qubits to control the order with which they traverse the gates \cite{Araujo2014,Nicolai2014,radu}.}
These could be any degrees of freedom of the physical system that the qubits are encoded in.
For example, trapped ions possess many electronic and vibrational modes, many of which could be suitably controlled \cite{Nicolai2014}.
In our experiment, we use a spatial degree of freedom of photonic qubits to create a superposition of different gate orders acting on a qubit encoded in the photon's polarization.
The use of multiple degrees of freedom of a single photon is a well-known tool in photonic quantum computing, which has allowed for many high-precision implementations of quantum information protocols \cite{Englert2001,lanyon2009, zhou2011,Rozema2014}.
Note that a different method for superimposing quantum gate orders was recently proposed for adiabatic quantum computing \cite{nakago2013parallelized}.

At the centre of our implementation is a Mach-Zehnder interferometer with a loop in each arm (see Fig. 2), which allows us to create the required superposition of gate orders.
In particular, it enables one qubit (qubit 1 of the circuit in Fig. 1d), encoded in a spatial degree of freedom of the photon, to coherently control the order in which two gates are applied to another qubit (qubit 2 of Fig. 1d), encoded in the photon's polarization.
Briefly, a single photon is sent to a 50/50 beamsplitter, which creates the spatial qubit: $\ket{0}$ if transmitted, and $\ket{1}$ if reflected.
The unitary gates $U_1$ and $U_2$ are implemented on the polarization state of the same photon using a set of waveplates.
Now, dependent on whether the photon is reflected or transmitted, the polarization qubit will see either $U_1U_2$ or $U_2U_1$.
{ The} two paths then coherently recombine on a final 50/50 beamsplitter (enacting the Hadamard gate shown in Fig. 1c).
Finally, simply measuring the state of the spatial qubit (i.e. whether the photon exits port 0 or 1 of the final beamsplitter) tells us if $U_1$ and $U_2$ commute or anti-commute.

\begin{figure} [t]
\begin{center} 
\scalebox{.56}{\includegraphics{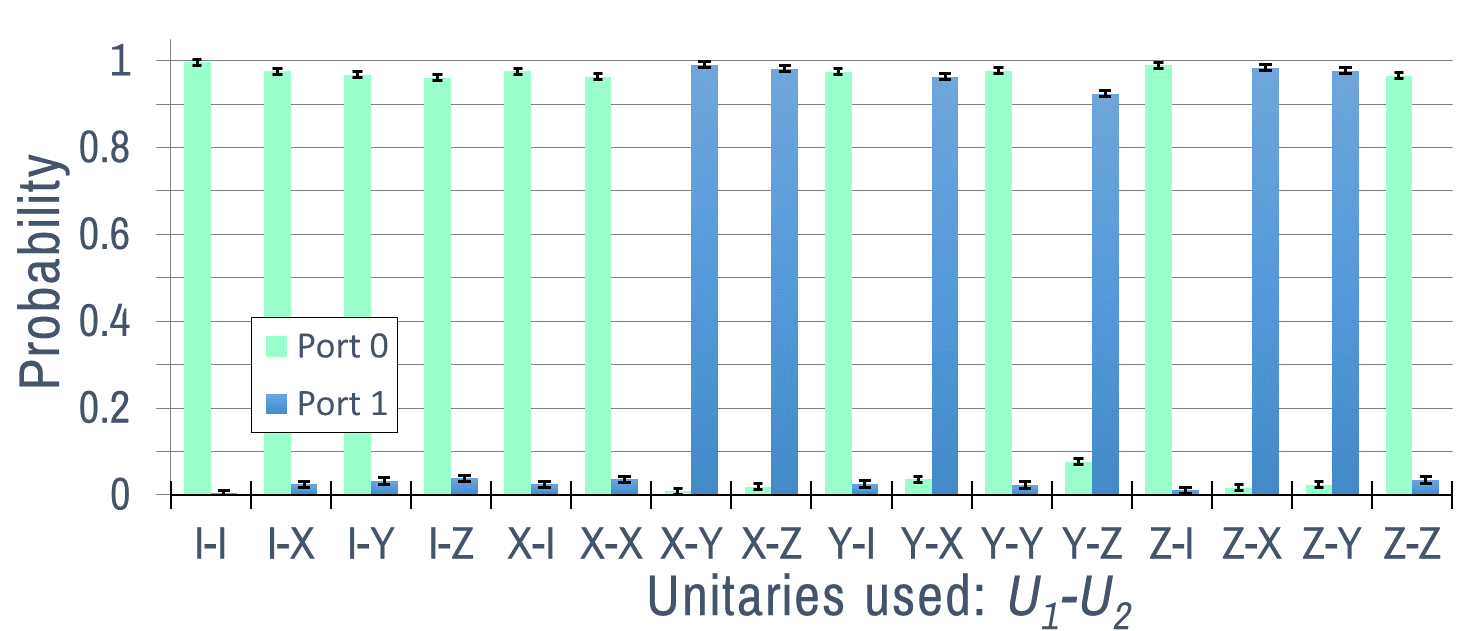}}
\caption{\footnotesize 
Experimental data showing { the probability with} which the photon {exits from a port }  when determining if a pair of random gates commute or anti-commute.
The blue bars are the experimentally observed probabilities for the photon to exit port 1, and the green bars to exit port 0.
If the gates commute, then, ideally, the photon should always exit port 0, while if they anti-commute the photon should exit port 1.
The x-axis is labelled with the choice of $U_1$ and $U_2$, where I is identity, X$=\sigma_x$, Y$=\sigma_y$, and Z$=\sigma_z$.
The average success rate (probability to exit the ``correct port'') of these data is $0.973\pm0.016$.}
\label{pauliData}
\end{center}
\end{figure}

To verify the successful implementation of this protocol we tested its performance on a number of representative unitary gates.
The first set of gates we tested were the four Pauli gates (including identity), $\{ \sigma_x,\sigma_y,\sigma_z,\mathcal{I}\}$.
The Pauli gates have simple commutation and anti-commutation relationships:
each gate only commutes with itself and identity, and anti-commutes with the other gates. 
For example, $\sigma_x$ commutes with $\mathcal{I}$ and $\sigma_x$, and anti-commutes with $\sigma_y$ and $\sigma_z$.
Thus, setting $U_1 = \sigma_x$ means that when $U_2$ is set to either $\mathcal{I}$ or $\sigma_x$ the photon will always exit port 0.
On the other hand, if $U_2$ is set to $\sigma_y$ or $\sigma_z$, the photon should always exit port 1.

\begin{figure}
\begin{center} 
\scalebox{.56}{\includegraphics{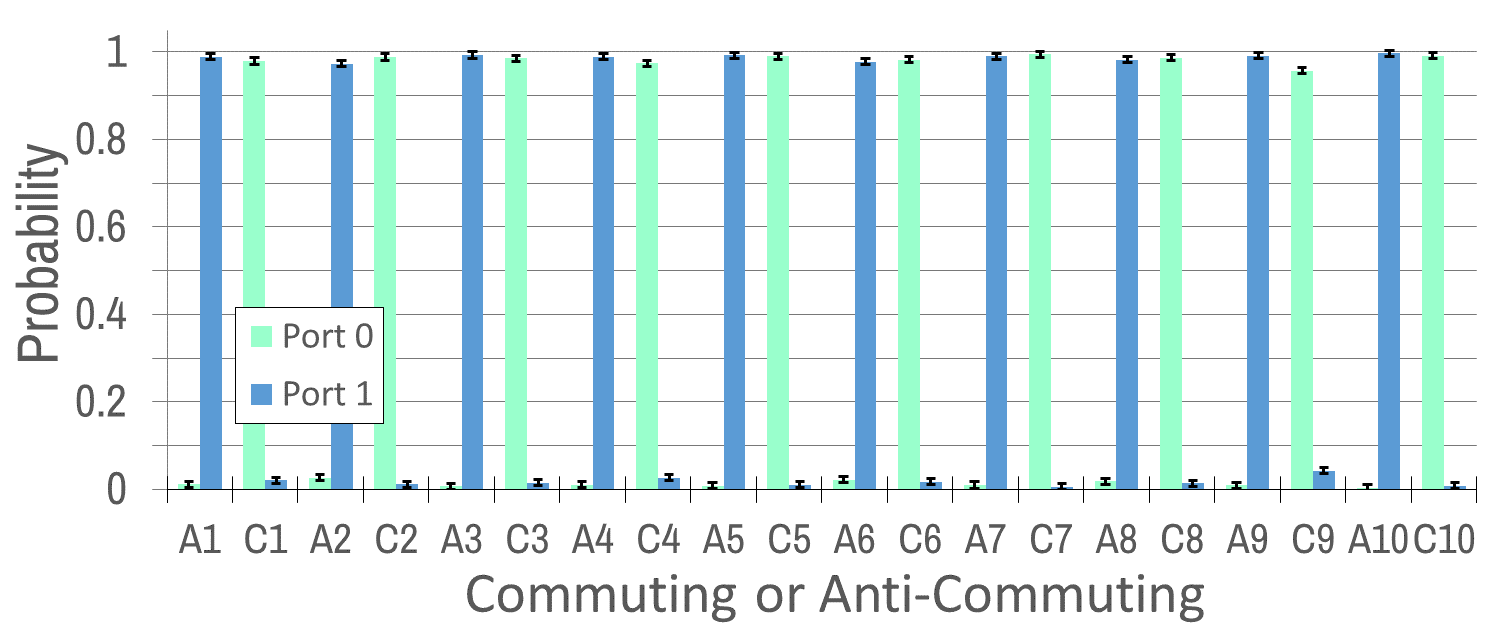}}
\caption{\footnotesize 
Experimental data showing { the probability with} which the photon {exits from a port }  when determining if a pair of random gates commute or anti-commute.
50 commuting and 50 anti-commuting pairs of gates were tested, {of which 10 for} each case are shown here.
The full data set is presented in the Appendix.
{ The data representation in this figure follows the same convention as in Fig.\ 3.}
However, here the x-axis is labelled A$i$ for anti-commuting case number $i$, and C$i$ for commuting case number $i$.
} \label{randData}
\end{center}
\end{figure}

To acquire data, $U_1$ and $U_2$ were first set to identity so that the phase of the interferometer could be set to { $\pi$; this was done using a piezo-driven mirror (PDM in Fig. 2)}.
Then $U_1$ and $U_2$ could be set to any desired single-qubit unitary gate by setting the waveplate angles appropriately.
For the Pauli-gate data, we cycled $U_1$ and $U_2$ through all 16 possible permutations and monitored the photon counts out of each port of the final beamsplitter.
For every Pauli-gate combination, the probability for the photons to exit each port was estimated (see the Appendix, Section A for details, including a discussion of the error bars).
The resulting probabilities are plotted in Fig. 3.
When the gates commute, we expect all of the photons to exit port 0, while if they anti-commute they should all exit port 1.
Our observed data agrees very well with this prediction.
For the Pauli gates, we were able to successfully determine whether a pair of gates commuted or anti-commuted with a success rate (probability to exit the correct port) of $0.973\pm0.016$.
For this data, the { initial} polarization state was $(\ket{H}+\ket{V})/\sqrt{2}$, but, as we verified experimentally, the protocol is independent the polarization (see the Appendix, Fig. 6).

Without using a superposition of different quantum circuits, it is impossible to perfectly determine if two gates commute or anti-commute with a single use of each gate\cite{Chiribella2012}.
{However, using a fixed-order quantum circuit, this task can be accomplished with an average success rate of at most 0.9288. The calculation, based on the ``quantum comb'' formalism \cite{chiribella08,araujo2015,gutoski2007toward,chiribella2012optimal}, is presented in the Appendix. }
In order to rigorously compare our protocol to such a quantum circuit, we randomly generated 50 pairs { of} commuting gates and 50 pairs { of} anti-commuting gates (see the Appendix, Section A) and we tested our protocol with each pair.
A subset of the data for these gates is plotted in Fig. 4.
The success rate of our protocol over the 100 pairs of commuting and anti-commuting gates is $0.976\pm0.015$, which surpasses the fixed-order bound by more than three standard deviations.

From a foundational point of view, our experiment can be seen as the first realization of a ``superposition of causal orders'', which represents an instance of an indefinite causal structure \cite{Oreshkov2012}. 
More generally, this work highlights the role of causal structures in quantum mechanics, a topic that has recently received an increasing theoretical \cite{Hardy,rudolph2012quantum,wood2012lesson} and experimental \cite{fitzsimons2013quantum,ried2014inferring} attention.
Here we have accomplished a task which is impossible if quantum operations are performed in any fixed, definite causal order. 
Our demonstration of superimposed quantum circuits illustrates that removing the requirement of a fixed gate order can provide quantum algorithms with real practical advantages.

%\newpage
\noindent\textbf{Acknowledgements}\\
We thank T. Batalh\~{a}o, A. Feix, C. Greganti, F. Massa for stimulating discussions, and T. R\"ogelsperger and J. Thomson for assisting with the figures.\\
We acknowledge support from the European Commission, QUILMI (No. 295293), EQUAM (No. 323714), PICQUE (No. 608062), GRASP (No.613024), QUCHIP (No.641039), and RAQUEL (No. 323970); 
the Austrian Science Fund (FWF) through START (Y585-N20), the doctoral programme CoQuS, and Individual Project (No. 2462);
the Vienna Science and Technology Fund (WWTF, grant ICT12-041);
the United States Air Force Office of Scientific Research (FA8655-11-1-3004);
the John Templeton Foundation;
and the Foundational Questions Institute (FQXi).
L.A.R was partially funded by a Natural Sciences and Engineering Research Council of Canada (NSERC) Postdoctoral Fellowship, and E.G.D. was partially funded by the Herchel Smith - Harvard Undergraduate Research Fellowship.\\

%%%%%%%%%%%%%%%%%%%%%%%%%%%%%%%%%%%%%%%%%%%%%%%%%%%%%%%%%%%%%%%%%%%%%%%
\bibliography{Superposition_paper_5} 
%\newpage
\cleardoublepage
\section*{Appendix}

\subsection{Experimental Details}
\noindent\textbf{Single-Photon Source --} 
The source generated photon pairs, in a separable polarization state, by means of the process of spontaneous parametric down conversion using a Sagnac loop \cite{TKim}. 
The Sagnac loop was built using a dual-wavelength polarizing beamsplitter (dPBS) and two mirrors. 
A type-II collinear periodically-poled Potassium Titanyl
Phosphate (PPKTP) crystal of length 20 mm was placed inside the loop and pumped by a 23.7 mW diode laser centred at 395 nm. 
Photon pairs were created at degenerate wavelength 790 nm. 
We set the pump beam polarization to be horizontal in order to generate the down-converted photons in a separable polarization state $\ket{H}\ket{V}$. 
The dichroic mirror (DM) transmited the pump beam and reflected the down-converted photons, and the half waveplate (HWP) and quarter waveplate (QWP) were used to adjust the polarization of the pump beam. 
Long (LP) and narrow band (BP) pass filters blocked the pump beam and selected the desired down-converted wavelength. 
Polarizers were aligned to transmit only down-converted photons with the desired polarization. 
After this, the down-converted photon pairs were coupled into single-mode fibres (SMF), and one photon from the pair was used as a herald while the other single photon was sent to our interferometer using a fibre collimator (FC). 

\noindent\textbf{Optical Implementation --}
We implemented our protocol using a Mach-Zehnder interferometer with two symmetric arm loops, see Fig. (\ref{expt}). 
After the first BS, the reflected and transmitted beams were sent to a different combination of waveplates which formed the unitary gates. 
The symmetric loops were built to have the same input direction through the unitary gates for both the transmitted and reflected beams; 
i.e. the photons always traversed the waveplates in the same direction, regardless of whether they saw $U_1$ or $U_2$ first.
After this, the two paths were recombined on the second BS. 
For the reflected beam, a HWP at $0^\circ$ was used before the waveplates implementing $U_2$ to ensure that the polarization state of the reflected and transmitted paths was the same before the unitary gates; this was required because the reflected path picked up a reflection phase not seen by the transmitted mode.
Another a HWP at $0^\circ$ was used in the transmitted arm after $U_2$ in order to compensate { the reflection from the second BS.}

After the interferometer, the photons exiting port 0 and 1 were coupled into single-mode fibers.
Then, the single photons were detected using avalanche photodiodes (APDs) which were connected to a home-built coincidence counter based on Spartan 3E FPGA to register two-photon events between either port and the heralding photon.
Since the coupling efficiency of each port and the detection efficiency of each APD were slightly different we had to correct for this to calculate the probabilities reported in figures 3 and 4.
To perform this correction, we varied the phase of the interferometer to send all of the photons from port 0 to port 1, and we recorded the counts out of port 0 ($C_0$) and the counts out of port 1 ($C_1$) as the phase was varied.
We then computed an efficiency factor $\eta$, such that $C_0 + C_1/\eta$ was constant.
{ We found that $\eta$ was typically around $0.7$, but its exact valued varied because the coupling efficiencies changed slightly from one day to the next.}
Using this, we estimated the probability to exit port 0 as $P_0=C_0/(C_0 + C_1/\eta)$, and the probability to exit port 1 as $P_1=1-P_0$.

The visibility of the interferometer was 99.4 $\pm$ 0.2 with a phase drift less than 9 mrad per minute. 
A complete data set (20 waveplate settings) was acquired in about 2 minutes (1 second of data was taken at each setting, so the majority of the time was spent moving waveplates).
The observed phase drift would lead to a negligible error (of only $\approx 0.02 \%$) over the 2 minute measurement time.
At each setting approximately 40,000 photon pairs were observed.
The error bars were estimated by performing each measurement 5 times and observing the standard deviation.
Each measurement setting had a slightly different standard deviation, but for convenience we took the largest standard deviation as the error bar for each measurement setting.
The dominant contribution to these fluctuations was a phase drift caused by rotating the waveplates, and the Poissonian error bars (due to 40,000 counts) are much smaller than these observed fluctuations.

\noindent\textbf{Polarization Unitary Gates --}
It is well known that the combination of three waveplates (in a quarter-half-quarter configuration) can implement an arbitrary single-qubit polarization gate.
Since this method is completely general, we used it to implement each of the two unitary gates $U_1$ and $U_2$.
Each of the six waveplates were mounted using a motorized rotation mount, which allowed the unitary gates to be set remotely while only minimally disturbing the phase of the interferometer. 
However, we still found a slight systematic phase drift when the waveplates were rotated.
We attributed this to slightly ``wedged'' shaped waveplates, which could change the optical path of each interferometer arm differently.
For our waveplates (true-zero order waveplates from Special Optics, which were the most parallel  waveplates we tested), we observed a maximum phase drift of 0.002 radians per degree of waveplate rotation.
The waveplate angles used to implement the different polarization gates are tabulated in Tables I and II.
On average, setting the six waveplates to implement a specific $U_1$ and $U_2$ required a total rotation of $\approx 45^\circ$; this would introduce a systematic error in the phase of the interferometer of about $0.009$ radians and an additional error onto our measured probabilities of $0.4\%$.

\noindent\textbf{Generating Commuting and Anti-Commuting Pairs of Gates --}
We tested our protocol on 50 randomly chosen pairs of commuting gates and on 50 randomly chosen pairs { of} anti-commuting gates.
To generate these gates we followed a very simple protocol.
First, we randomly chose a unitary gate $\mathcal{R}$ from the Haar measure.
Next, to generate two anti-commuting gates $A_1$ and $A_2$, we set $A_1 = \mathcal{R}\sigma_z\mathcal{R}^\dagger$ and $A_2 = \mathcal{R}\sigma_y\mathcal{R}^\dagger$ (where $\sigma_y$ and $\sigma_z$ are Pauli gates).
Then, two commuting gates were defined from $\mathcal{R}$ as 
$C_1 = \mathcal{R} \left(\begin{array}{cc} 1 & 0\\0 & e^{i\theta_1} \end{array}\right) \mathcal{R}^\dagger$
and
$C_2 = \mathcal{R} \left(\begin{array}{cc} 1 & 0\\0 & e^{i\theta_2} \end{array}\right) \mathcal{R}^\dagger$,
where $\theta_1$ and $\theta_2$ where also chosen randomly (we selected them from a uniform distribution between $0$ and $2\pi$).
After this we computed the waveplate angles required to implement $C_1$, $C_2$, $A_1$, and $A_2$ (these angles are listed in Table II).

\subsection{A Fixed-Order Quantum Circuit for Our Task}

In Ref. 3 it was proven that quantum circuits with a fixed
gate order (fixed-order circuits) cannot distinguish between commuting and
anti-commuting unitary gates with probability one under the condition that each unitary is queried once. Theoretically, exploiting a
superposition of different gate orders allows for the two cases to be
distinguished perfectly under the same condition. However, since an experimental implementation of the
superposition necessary involved errors, one should compare our experiment to a fixed-order
circuit which can probabilistically distinguish between the
two cases. Here we will show that our experimental success probability is higher
than the maximal probability of success of any fixed-order circuit.

We will assume that to determine if two gates commute or anti-commute we will
 run a fixed-order circuit  that contains a single copy of  $U_1$ and $U_2$ and afterwards measure a single qubit. If the result of this single-qubit
measurement is zero (one) we will say the two gates commute (anticommute). Then,
the probability of success is defined as the average of the probability of
measuring zero given that the unitary gates anti-commute and the probability of
measuring one given that the unitary gates commute. In symbols:

\begin{equation}
p_\text{succ} = \frac12 (p(0|C) + p(1|A)).
\end{equation}

\begin{figure}
\begin{center} 
\scalebox{.63}{\includegraphics{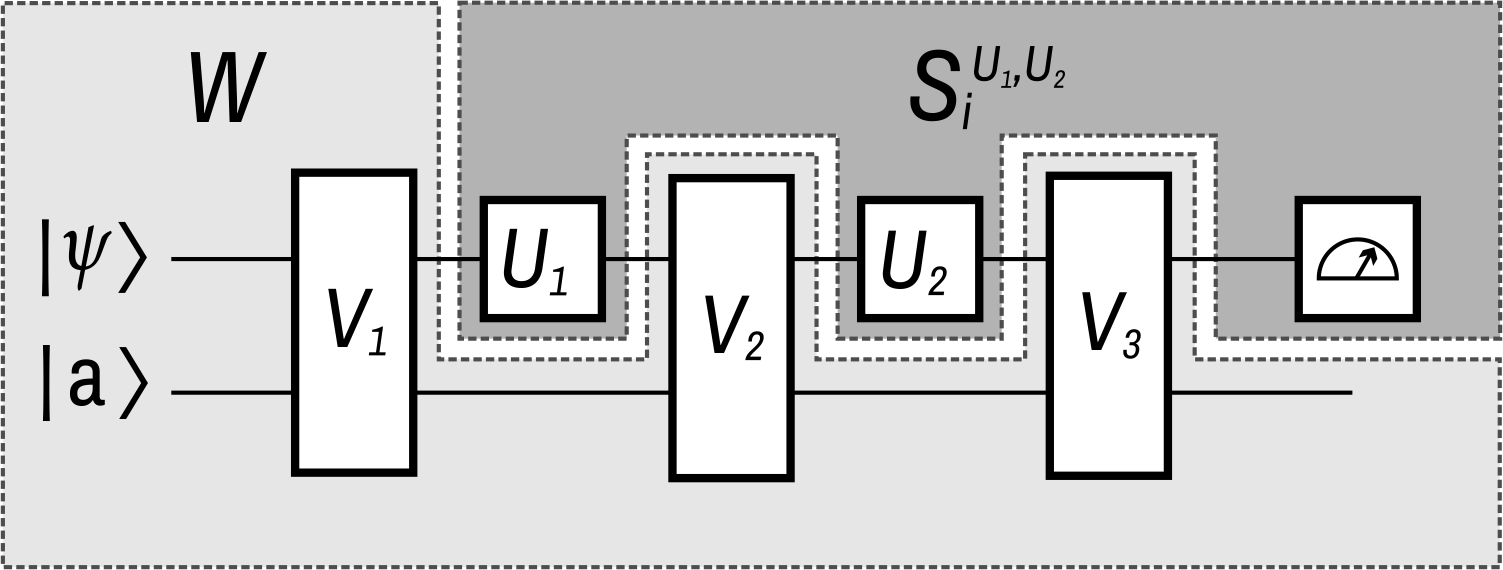}}
\caption{\footnotesize Circuit representation of the operators $W$ and $S_i^{U_1,U_2}$.  Here $\ket{a}$ is an ancilla of arbitrary dimension, and $V_1$,$V_2$, and $V_3$ are arbitrary unitary gates acting on the system and the ancilla.} \label{w}
\end{center}
\end{figure}

To calculate $p_\text{succ}$, we use tools from  Ref. [21-24].
There it is shown that one can separate any circuit that is applied to determine if
the unitaries commute or anticommute in two parts: the first is an operator

\begin{equation}
S_i^{U_1,U_2} = \mathfrak{C}(U_1) \otimes \mathfrak{C}(U_2) \otimes
\ket{i}\bra{i}, \hspace{2ex} i=0,1
\end{equation}
 
\noindent that represent the unitaries $U_1$ and $U_2$ and the outcome $i$ of a measurement in the $Z$
basis, where $\mathfrak{C}(U_j)$, with $j=1,2$, is the Choi-Jamio{\l}kowski operator of the
unitary $U_i$. The second is an operator $W$ that represents the most general
fixed-order circuit (or even a convex combination of fixed-order circuits) that
connects the unitaries and the measurement.
A circuit representation of these operators is shown in Fig. \ref{w}.
In this case,

\begin{equation}
p(i|U_1,U_2) = \tr (S_i^{U_1,U_2} W)
\end{equation}

\noindent is the probability of obtaining outcome $i$ after applying unitaries $U_1$ and $U_2$.

The probabilities $p(0|C)$ and $p(1|A)$ are then the averages

\begin{align}
p(0|C) &= \int \mathrm{d}\mu_C p(0|U_1,U_2) \nonumber\\ 
&=\tr \left[\left( \int \mathrm{d}\mu_C S_0^{U_1,U_2}\right) W\right]. 
\end{align}

\begin{align} 
p(1|A) &= \int \mathrm{d}\mu_A p(1|U_1,U_2) \nonumber \\
&= \tr \left[\left( \int \mathrm{d}\mu_A S_1^{U_1,U_2}\right) W\right].
\end{align}

\noindent where $\mathrm{d}\mu_C, \mathrm{d}\mu_A$ are measures on the set of
commuting and anticommuting pairs of unitaries, respectively.  One should note,
now, that $p_\text{succ}$ depends crucially on the choice of measures
$\mathrm{d}\mu_C, \mathrm{d}\mu_A$. For instance, if one selects $U_1$ and $U_2$
only from the Pauli matrices, then $p_\text{succ} = 1$. We chose measures, shown
in the Section A, that  can generate any pair of commuting or anticommuting unitaries, modulo a global phase. Defining the operators

\begin{eqnarray}
 S_0^C=  \int \mathrm{d}\mu_C S_0^{U_1,U_2} \\
 S_1^A = \int \mathrm{d}\mu_A S_1^{U_1,U_2},
\end{eqnarray}

\noindent we have finally that the probability of success can be expressed as

\begin{equation}
p_\text{succ} = \tr[ W\left(S_0^C + S_1^A\right)/2], 
\end{equation}

\noindent which is a linear function of $W$. Since one can characterize the set of
fixed-order circuits through linear constraints on positive semidefinite
operators, optimizing $p_\text{succ}$ is a semidefinite program (SDP), which can
be guaranteed to reach the global optimum [22-24]. Solving this
optimization problem numerically, we found that

\begin{equation}
p_\text{succ} = 0.9288, 
\end{equation}

\noindent which is significantly lower than the success probability of $0.976\pm0.015$
that we measured experimentally.\\

As an extra verification, we tested the specific 100 pairs of gates that we used in our experiment with the optimal fixed-order circuit.
We found that this optimal fixed-order circuit has success rate of $0.9390$ averaged on these 100 pairs of gates -- still much lower than our experimental success probability.

\subsection{Additional Raw Data}

As we discussed in the main text, our protocol to determine if $U_1$ and $U_2$ commute or anti-commute is independent of the state of qubit 2 (defined in Fig. 1d and Eq. 1 of the main text).
In our experiment, qubit 2 is encoded in the polarization state of our single photons. 
As an additional verification of our implementation we also tested the state dependence of our protocol experimentally.
To do this, we simply changed the state of qubit 2, by rotating the state-preparation half waveplate, and repeated our protocol for several different input states.
We rotated this HWP in steps of $10^\circ$, from $0^\circ$ to $40^\circ$ and took ``Pauli-gate data'' (described in the main text) for each waveplate setting.
This data is plotted in Fig. 2.
There were no significant changes to the performance of our protocol for any of these settings, and the success rate averaged over these five states was $0.970\pm 0.024$.

In addition to the data presented in the main text, we verified our protocol's operation on 50 different pairs of commuting gates and 50 different pairs of anti-commuting gates.
The 50 different cases were taken in groups of 10, and the data for each group was acquired in the same way as the Pauli gates.
Before acquiring each group data, the phase of the interferometer was set to $\pi$ while $U_1$ and $U_2$ were set to implement identity.
Then $U_1$ and $U_2$ were set to implement some other gates, and photons were counted for $1$ second on each setting.
These data are presented in Fig. 3, and they exhibit an average success rate of $0.976\pm 0.015$.

\subsection{Waveplate Settings}

\begin{table}[t]
\centering
\begin{tabular}{|c|c|c|c|c|c|c|c|}
\hline
&
\multicolumn{3}{c|}{Angles for $U_1$}&\multicolumn{3}{c|}{Angles for $U_2$}\\
$U$ & Q1 & H1 & Q2 & Q3 & H2 & Q4 \\
\hline
%FIRST ROW
$\mathcal{I}$ 
&0.0	&0.0	&0.0	&0.0	&0.0	&0.0\\
%SECOND ROW
$\sigma_x$ 
&0.0	&45.0	&0.0	&0.0	&45.0	&0.0\\
%THIRD ROW
$\sigma_y$
&90.0	&45.0	&0.0	&45.0	&0.0	&-45.0\\
%FOURTH ROW
$\sigma_z$
&90.0	&0.0	&0.0	&0.0	&0.0	&90.0\\
\hline
\end{tabular}
\caption{The waveplate angles used to implement the four Pauli gates. The first column lists the desired unitary gate, the next three columns list the waveplate angles (quarter waveplate, half waveplate, then quarter waveplate) that were used to implement the gate for $U_1$, and the final three columns list the waveplate angles that were used for $U_2$.}
\end{table}

Each unitary gate that we tested was implemented with three waveplates: first a quarter waveplate, then a half waveplate, followed by a final quarter waveplate.
Thus we need a total of six waveplates for two unitary gates.
Each waveplate was also mounted in a motorized rotation mount, allowing it to be remotely set without disturbing the phase of the interferometer.

For completeness, here we list the waveplate settings used to implement the various unitary gates that we tested.
Table I lists the waveplate angles for the four Pauli gates.
For experimental convenience, we used different angles to set $U_1$ and $U_2$ to $\sigma_y$ and $\sigma_z$.
This has no effect on the logical gate that the waveplates apply.
In fact, this can be seen as a further verification of our protocol: the protocol's success is independent of the physical apparatus used to enact each gate.

The next set of 100 gates that we tested were randomly generated commuting and anti-commuting gates.
As described in the Section A, we first randomly generate a gate $\mathcal{R}$.
Two anti-commuting gates $A_1$ and $A_2$ were generated from $\mathcal{R}$ by setting $A_1 = \mathcal{R}\sigma_z\mathcal{R}^{\dagger}$ and $A_2 = \mathcal{R}\sigma_y\mathcal{R}^{\dagger}$ (where $\sigma_y$ and $\sigma_z$ are Pauli gates).
Two commuting gates were constructed as 
$C_1 = \mathcal{R} \left(\begin{array}{cc} 1 & 0\\0 & e^{i\theta_1} \end{array}\right) \mathcal{R}^\dagger$
and
$C_2 = \mathcal{R} \left(\begin{array}{cc} 1 & 0\\0 & e^{i\theta_2} \end{array}\right) \mathcal{R}^\dagger$,
where $\theta_1$ and $\theta_2$ are chosen randomly between $0$ and $2\pi$ from a uniform distribution.
The waveplate angles that we used to implement these 100 pairs of gates are tabulated in Table II.

\newpage
\begin{figure*}[t]
\begin{center} 
\includegraphics[width = .85\textwidth]{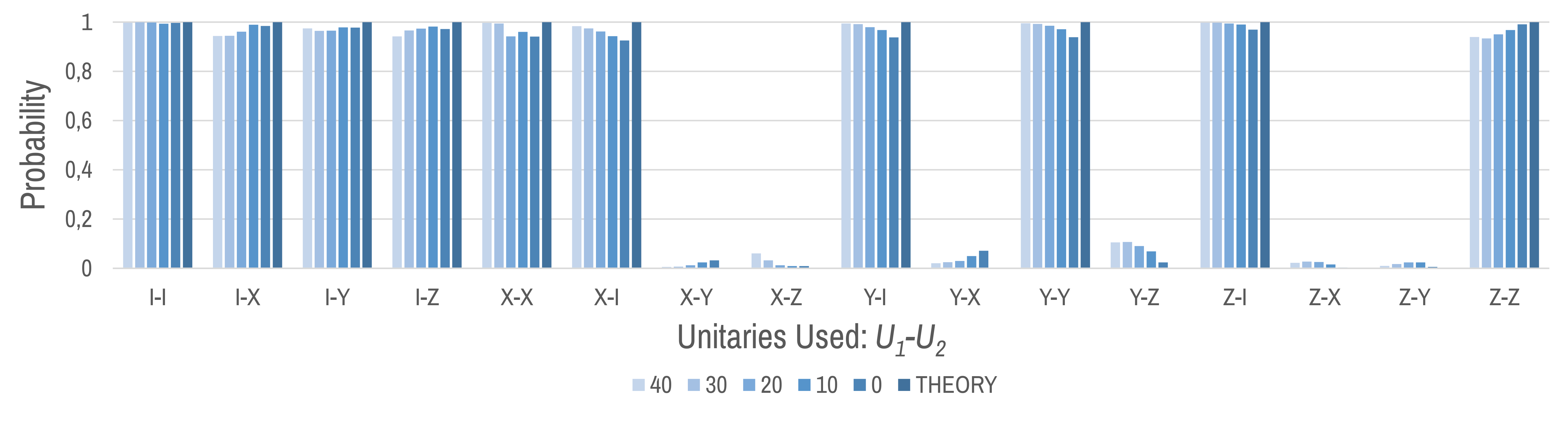}
\caption{\footnotesize Experimental data for determining if two Pauli gates commute or anti-commute for several different input states.
As in Fig. 3 of the main text, the x-axis is labelled with the choice of $U_1$ and $U_2$ (where I$=\mathcal{I}$ is the identity, X$=\sigma_x$, Y$=\sigma_y$, and Z$=\sigma_z$) and the y-axis is the probability for the photon to exit port 0. If the photon exits port 0, $U_1$ and $U_2$ commute, and if it exits port 1 they anti-commute. For clarity only the port 0 probabilities are shown here. The bars of different shades are the data for different input states; the legend lists the angle of the state-preparation half waveplate, which was used to set the input state.
The average success rate of these data is $0.970\pm 0.024$.
} \label{allStates}
\end{center}
\end{figure*}

\newpage
\begin{figure*}[t]
\begin{center} 
\includegraphics[width = 0.7\textwidth]{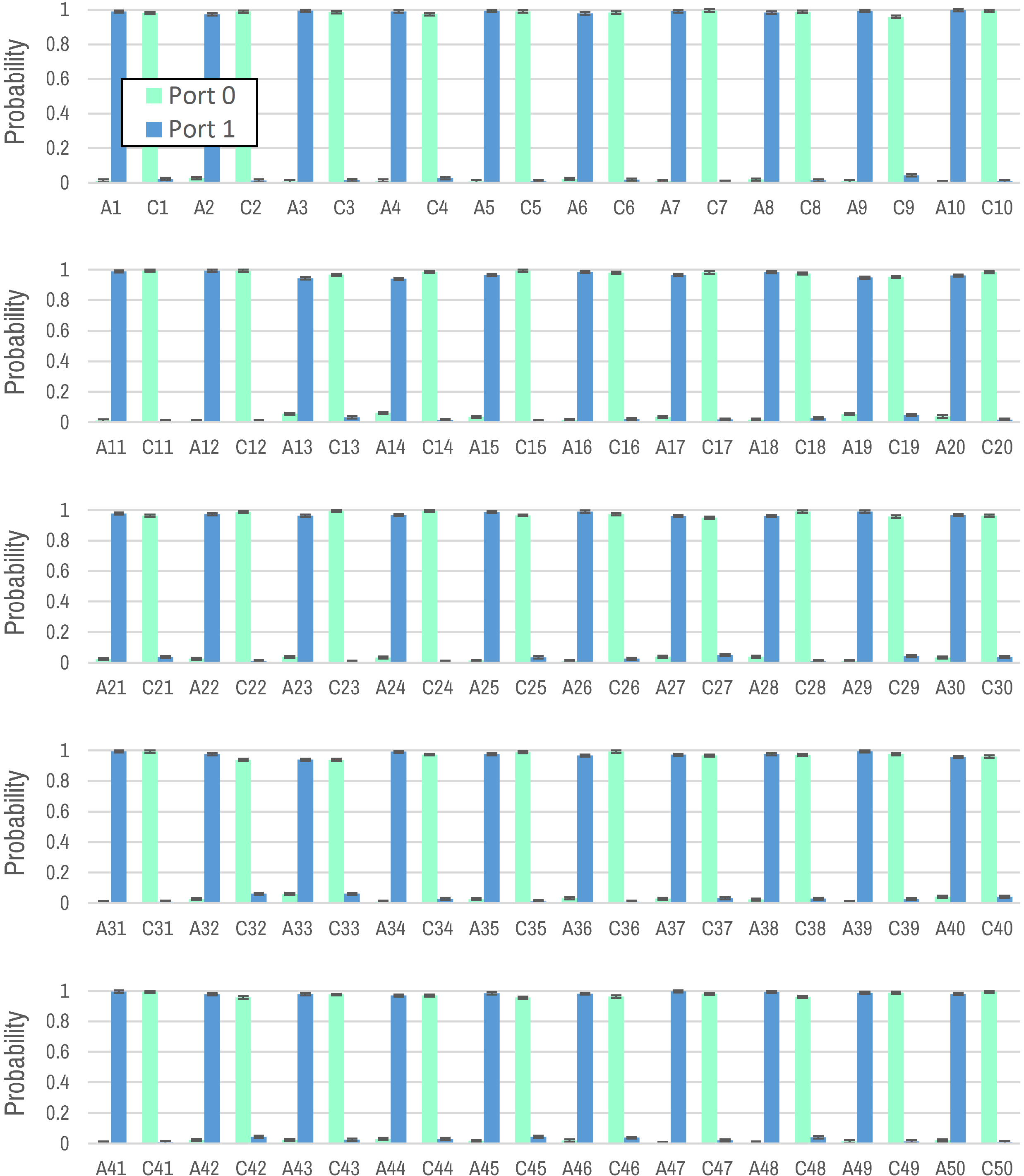}
\caption{\footnotesize Experimental data for determining if two random gates commute or anti-commute for 100 different pairs of unitary gates.
As in Fig. 4 of the main text, the x-axis is labelled with A$i$ for anti-commuting case number $i$, and C$i$ for commuting case number $i$.
For clarity, only one port is shown here.
The average success rate of these data is $0.976\pm 0.015$.
} \label{allUs}
\end{center}
\end{figure*}
\newpage

\begin{table*}[tbp]
%\tiny
\centering
\begin{tabular}{|c|c|c|c|c|c|c|c|c|c|c|c|c|c|}
\hline
&
\multicolumn{6}{c|}{Commuting Angles}&\multicolumn{6}{c|}{Anti-Commuting Angles}\\
&
\multicolumn{3}{c|}{$C_1$}&\multicolumn{3}{c|}{$C_2$}&\multicolumn{3}{c|}{$A_1$}&\multicolumn{3}{c|}{$A_2$}\\
Index & Q1 & H1 & Q2 & Q3 & H2 & Q4 & Q1 & H1 & Q2 & Q3 & H2 & Q4\\
\hline
1 & 25.61 & 5.20 & 24.38 & 24.97 & 25.80 & 25.02 & 69.99 & 70.74 & -20.01 & 110.81 & 70.89 & 20.81 \\
2 & 25.64 & 23.90 & 53.26 & 47.93 & 49.85 & 30.96 & 84.45 & 65.67 & -5.55 & 61.93 & 83.23 & -28.07 \\
3 & 79.12 & 47.13 & 39.88 & 107.12 & 41.53 & 11.89 & 104.50 & 41.47 & 14.50 & 89.83 & 73.62 & -0.17 \\
4 & 2.27 & 8.62 & 6.33 & -17.71 & 32.37 & 26.31 & 49.30 & 61.80 & -40.70 & 100.93 & 114.02 & 10.93 \\
5 & 106.91 & 26.73 & -15.24 & 42.54 & 48.87 & 49.12 & 90.83 & 24.37 & 0.83 & 62.14 & 48.83 & -27.86 \\
6 & 138.45 & 73.49 & -30.11 & 49.73 & 38.22 & 58.61 & 99.17 & 182.20 & 9.17 & 88.31 & 35.83 & -1.69 \\
7 & 30.10 & 34.04 & 23.82 & 40.93 & 50.55 & 12.99 & 71.96 & 60.23 & -18.04 & 91.00 & 121.56 & 1.00 \\
8 & 69.12 & 32.05 & 25.94 & 21.59 & 64.82 & 73.47 & 92.53 & 26.91 & 2.53 & 88.40 & 67.93 & -1.60 \\
9 & 17.17 & 29.40 & 4.78 & 10.68 & 9.93 & 11.28 & 55.98 & 47.99 & -34.02 & 91.23 & 115.91 & 1.23 \\
10 & 66.71 & 29.40 & 20.31 & 107.06 & 28.21 & -20.04 & 88.51 & 25.23 & -1.49 & 53.48 & 46.39 & -36.52 \\
11 & 118.60 & 105.98 & -34.42 & 35.79 & 148.00 & 48.39 & 87.09 & 6.78 & -2.91 & 107.73 & 75.27 & 17.73 \\
12 & 13.17 & 14.76 & 30.13 & 78.31 & 40.49 & -35.02 & 66.65 & 41.68 & -23.35 & 67.14 & 87.17 & -22.86 \\
13 & 29.19 & 51.62 & 15.89 & 19.52 & 4.35 & 25.56 & 67.54 & 63.47 & -22.46 & 108.32 & 41.21 & 18.32 \\
14 & 39.98 & 34.80 & 36.36 & 39.66 & 35.38 & 36.67 & 83.17 & 7.23 & -6.83 & 49.31 & 31.63 & -40.69 \\
15 & 31.87 & 44.00 & 41.97 & 31.16 & 44.93 & 42.68 & 81.92 & 9.33 & -8.08 & 74.76 & 47.59 & -15.24 \\
16 & -7.47 & 30.83 & 25.75 & -24.11 & 38.00 & 42.39 & 54.14 & 69.19 & -35.86 & 75.63 & 49.82 & -14.37 \\
17 & 47.80 & 81.82 & -2.10 & 22.27 & 24.28 & 23.43 & 67.85 & 78.89 & -22.15 & 65.59 & 31.66 & -24.41 \\
18 & 17.57 & 24.35 & 12.72 & 54.18 & 52.86 & -23.89 & 60.15 & 53.01 & -29.85 & 97.45 & 30.56 & 7.45 \\
19 & 18.30 & 145.88 & -53.14 & 1.14 & 61.19 & -35.98 & 117.58 & 90.00 & 27.58 & 78.13 & 81.95 & -11.87 \\
20 & 24.36 & 29.96 & 10.69 & 54.25 & 48.53 & -19.20 & 62.53 & 49.02 & -27.47 & 119.72 & 100.21 & 29.72 \\
21 & 81.68 & 4.23 & 38.15 & 148.16 & 66.28 & -28.34 & 104.91 & 187.37 & 14.91 & 48.04 & 19.93 & -41.96 \\
22 & -12.21 & -6.83 & -9.66 & -43.49 & 24.68 & 21.61 & 124.06 & 115.50 & 34.06 & 97.12 & 38.32 & 7.12 \\
23 & 23.49 & 27.69 & -38.97 & 167.49 & 68.11 & -2.96 & 37.26 & 28.70 & -52.74 & 68.18 & 6.62 & -21.82 \\
24 & 51.36 & 52.70 & 49.32 & 46.18 & 41.08 & 54.50 & 95.34 & 83.67 & 5.34 & 82.13 & 24.26 & -7.87 \\
25 & -11.32 & 79.67 & 25.49 & 51.83 & 31.71 & -37.66 & 52.08 & 31.71 & -37.92 & 83.82 & 7.52 & -6.18 \\
26 & -7.62 & 41.69 & 28.74 & 0.43 & 34.57 & 20.70 & 55.56 & 64.22 & -34.44 & 122.61 & 65.55 & 32.61 \\
27 & 28.05 & 6.10 & 11.74 & 52.45 & 80.22 & -12.66 & 64.89 & 79.02 & -25.11 & 35.26 & 103.48 & -54.74 \\
28 & 77.86 & 19.97 & 23.19 & 72.41 & 22.05 & 28.64 & 95.52 & 17.64 & 5.52 & 83.99 & 52.03 & -6.01 \\
29 & 58.96 & 81.60 & -17.09 & 23.84 & 16.07 & 18.03 & 65.93 & 81.22 & -24.07 & 60.13 & 30.68 & -29.87 \\
30 & 18.74 & 38.29 & 19.72 & 19.58 & 4.53 & 18.88 & 64.23 & 64.85 & -25.77 & 109.50 & 31.32 & 19.50 \\
31 & 48.00 & 37.76 & -13.85 & 21.01 & 20.97 & 13.14 & 62.08 & 39.56 & -27.92 & 59.53 & 81.95 & -30.47 \\
32 & 7.79 & 60.78 & -18.10 & 0.78 & 70.97 & -11.09 & 129.84 & 119.16 & 39.84 & 82.52 & 76.68 & -7.48 \\
33 & 100.79 & 84.36 & -37.71 & 41.77 & 0.53 & 21.32 & 76.54 & 81.80 & -13.46 & 29.01 & 41.72 & -60.99 \\
34 & -2.47 & -18.24 & -20.91 & -43.90 & 5.08 & 20.52 & 123.31 & 96.47 & 33.31 & 50.29 & 34.60 & -39.71 \\
35 & 70.25 & 59.09 & 35.26 & -40.49 & 54.02 & -34.00 & 97.75 & 63.46 & 7.75 & 108.68 & 118.68 & 18.68 \\
36 & 82.08 & 94.57 & 48.90 & 126.39 & 99.59 & 4.59 & 110.49 & 101.10 & 20.49 & 95.66 & 39.98 & 5.66 \\
37 & 59.51 & 57.89 & 41.49 & 33.06 & 37.51 & 67.94 & 95.50 & -19.28 & 5.50 & 50.65 & 50.77 & -39.35 \\
38 & 7.54 & 6.90 & 8.93 & -17.77 & 69.99 & 34.24 & 53.23 & 39.46 & -36.77 & 95.20 & 10.93 & 5.20 \\
39 & 8.81 & 57.10 & 37.25 & 18.95 & 41.33 & 27.11 & 68.03 & 73.44 & -21.97 & 103.03 & 72.63 & 13.03 \\
40 & 20.74 & 40.54 & 17.97 & 18.33 & 2.30 & 20.39 & 64.36 & 62.84 & -25.64 & 109.31 & 20.23 & 19.31 \\
41 & 58.10 & 32.61 & 40.43 & 50.68 & 46.22 & 47.85 & 94.27 & 106.66 & 4.27 & 35.45 & -31.98 & -54.55 \\
42 & 31.72 & 1.79 & 29.83 & 30.28 & 50.47 & 31.26 & 75.77 & 76.36 & -14.23 & 30.02 & 55.96 & -59.98 \\
43 & 22.65 & 31.45 & 28.14 & 40.86 & 0.87 & 9.93 & 70.40 & 82.42 & -19.60 & 62.93 & 30.32 & -27.07 \\
44 & 37.95 & 125.41 & 51.23 & 57.55 & 60.72 & 31.62 & 89.59 & 72.23 & -0.41 & 63.13 & 83.65 & -26.87 \\
45 & 39.99 & 58.07 & 14.13 & 26.84 & 26.10 & 27.28 & 72.06 & 65.53 & -17.94 & 84.90 & 32.69 & -5.10 \\
46 & 13.21 & 24.69 & 1.16 & 5.26 & 0.84 & 9.10 & 52.18 & 43.89 & -37.82 & 88.55 & 20.99 & -1.45 \\
47 & 80.68 & 89.06 & -29.71 & 1.99 & 48.40 & 48.97 & 70.48 & 88.17 & -19.52 & 100.80 & 83.49 & 10.80 \\
48 & 94.56 & 59.52 & -31.27 & 39.31 & 44.44 & 23.99 & 76.65 & 62.20 & -13.35 & 129.97 & 116.26 & 39.97 \\
49 & 30.87 & 36.27 & 60.40 & 52.65 & 50.37 & 38.62 & 90.64 & 62.90 & 0.64 & -51.99 & 43.14 & 38.01 \\
50 & 39.84 & 42.78 & 41.17 & 49.05 & 17.95 & 31.97 & 85.51 & 3.66 & -4.49 & 85.87 & 49.02 & -4.13 \\
\hline
\end{tabular}\caption{The waveplate angles used to implement the 100 randomly chosen commuting and anti-commuting gates.}
\end{table*}

\end{document}